\documentclass{phbauth}
\usepackage{graphicx}

\begin{document}

\begin{frontmatter}

\title{Various ordered states in a 2D interacting electron
system  close to an electronic topological transition}

\author[address1]{F. Bouis\thanksref{thank1}},
\author[address2]{M.N. Kiselev\thanksref{thank2}},
\author[address1]{F. Onufrieva},
\author[address1]{P. Pfeuty}

\address[address1]{Laboratoire L\'eon Brillouin, CE-Saclay, 91191 Gif-sur-Yvette, France}
\address[address2]{Russian Research Center "Kurchatov Institute",
123 182 Moscow, Russia}
\thanks[thank1]{Corresponding author. E-mail: bouis@llb.saclay.cea.fr. fax : 33-1-69088261}
\thanks[thank2]{Present address: Institut f. Theoret. Physik, Univ. W\"urzburg D-97074 W\"urzburg, Federal Republic of Germany}

\begin{abstract}
We consider a 2D electron system on a square lattice with hopping beyond
 nearest neighbors. The existence of the quantum critical point associated
 with an electronic topological
transition  in the noninteracting system results in density wave (DW)
 and high temperature d-wave superconducting (dSC) instabilities in the presence of an exchange interaction $J$.
 We analyse different DW ordering such as isotropic Spin DW (SDW), d-wave SDW,
  isotropic Charge DW (CDW) and d-wave CDW.
 The coexistence of dSC and SDW orders leads necessary to the existence of a third order
which is a $\pi$ triplet superconducting (PTS) order.
 A new phase diagram with a mixed phase of SDW, dSC and
 PTS order is found. The theory is applied to high-$T_c$ cuprates.
 \end{abstract}
\end{frontmatter}
A strong opinion exists in the physical community that the anomalous properties of high $T_c$ cuprates can reflect the existence of some hidden quantum critical
point (QCP). 
It has been suggested recently \cite{flo1} that such a QCP could be due to
 an electronic topological transition (ETT) in a 2D system. As was shown, the existence of this QCP results in DW and dSC instabilities of the normal state.
 In the present paper we study DW states of different nature and the mixed state with both DW and dSC ordering.

The starting point is a 2D model of interacting fermions on a square lattice
with hopping between nearest ($t$) and next nearest ($t'$) neighbors (while $t'/t<0$
that corresponds to the situation experimentally observed \cite{arpes}) :
%\begin{eqnarray}
%g_{SDW}  =  2U+2J^z &\mbox{~~~} & g_{dSDW}  =  b+J^z-2J^\perp
%\nonumber \\ g_{CDW}  =  -2U+2b  &\mbox{~~~} & g_{dCDW} =
%b+J^z+2J^\perp \nonumber
%\end{eqnarray}
\begin{eqnarray}
H & = & \sum_{\bf k} \epsilon_{\bf k} c^+_{{\bf k} \sigma} c_{{\bf k} \sigma} +\sum_i U n_{i
\uparrow} n_{i \downarrow} \nonumber \\  
  & \mbox{~} & +\sum_{\langle i,j \rangle} [J^z S_i^z
S_j^z +J^\perp S_i^+ S_j^- +\frac{b}{4} n_i n_j]   
\end{eqnarray} 
where $n_i= n_{i
\uparrow} + n_{i \downarrow}$ ($n_{i\sigma}=c^+_{i\sigma}c_{i\sigma}$) and ${\bf S}_i=c^+_{i\alpha}{\bf \sigma}_{\alpha\beta} c_{i\beta}$ are the charge and spin densities.
The Fermi Surface (FS) of noninteracting quasiparticles with the dispersion law $\epsilon_{\bf k}
 = -2t (\cos k_x + \cos k_y ) - 4t' \cos k_x \cos k_y$
 changes from close to open at the critical
doping $x_c$ resulting in the 2D ETT which is at the origin of two types of instabilities : dSC and DW.
Calculations show that the superconducting temperature is highest at $x_c$, thus the regimes
$x<x_c$  and $x>x_c$ could be seen respectively as underdoped and overdoped.  The shape of the FS in
the underdoped regime favors the DW
instabilities with the antiferromagnetic wave vector
${\bf Q}=(\pi,\pi)$. 
The DW order
parameter is $\psi_{\bf k}=\langle
c^+_{{\bf k+Q}\uparrow}c_{{\bf k}\uparrow} \pm c^+_{{\bf k+Q}\downarrow}c_{{\bf k}\downarrow}\rangle$. On one hand it could be of the spin $(-)$ or charge $(+)$
type and on the other hand it could be isotropic or have a certain dependence in
${\bf k}$. The ${\bf k}$-dependent case corresponds to the so-called unconventional DW order which was considered for example in \cite{ikeda} (see also \cite{comment}). Below we consider four different types of DW order : 
isotropic spin or charge density wave (SDW or CDW), d-wave spin or charge density wave (dSDW or dCDW. For all of them the spectrum is given by
\begin{equation}
E^\pm_{\bf k}=\frac{\epsilon_{\bf k} +\epsilon_{{\bf k}+{\bf Q}}}{2}\pm \sqrt{\frac{(\epsilon_{\bf k} -\epsilon_{{\bf k}+{\bf Q}}}{2})^2+\Delta_{\bf k}^2}
\end{equation}
with $\Delta_{\bf k}=\Delta_0$ for SDW and CDW orders, and
$\Delta_{\bf k}=\Delta_0[\cos k_x -\cos k_y]$ for dSDW and dCDW orders.
The effective coupling constants are given respectively by
\begin{eqnarray}
g_{SDW}  =  2U+2J^z &\mbox{~~~} & g_{dSDW}  =  b+J^z-2J^\perp
\nonumber \\ g_{CDW}  =  -2U+2b  &\mbox{~~~} & g_{dCDW} =
b+J^z+2J^\perp \nonumber
\end{eqnarray}
In the case of the $t-J$ model ($J^z=J^\perp=-b$ and
$U=0$) which we use below to describe the high $T_c$ cuprates, only two instabilities, dCDW and SDW, are possible, being determined by equal effective constants, $g_{SDW}=g_{dCDW}$. 

\begin{figure}[btp]
%h=here, t=top, b=bottom, p=separate figure page
\begin{center}\leavevmode
\includegraphics[width=1.\linewidth]{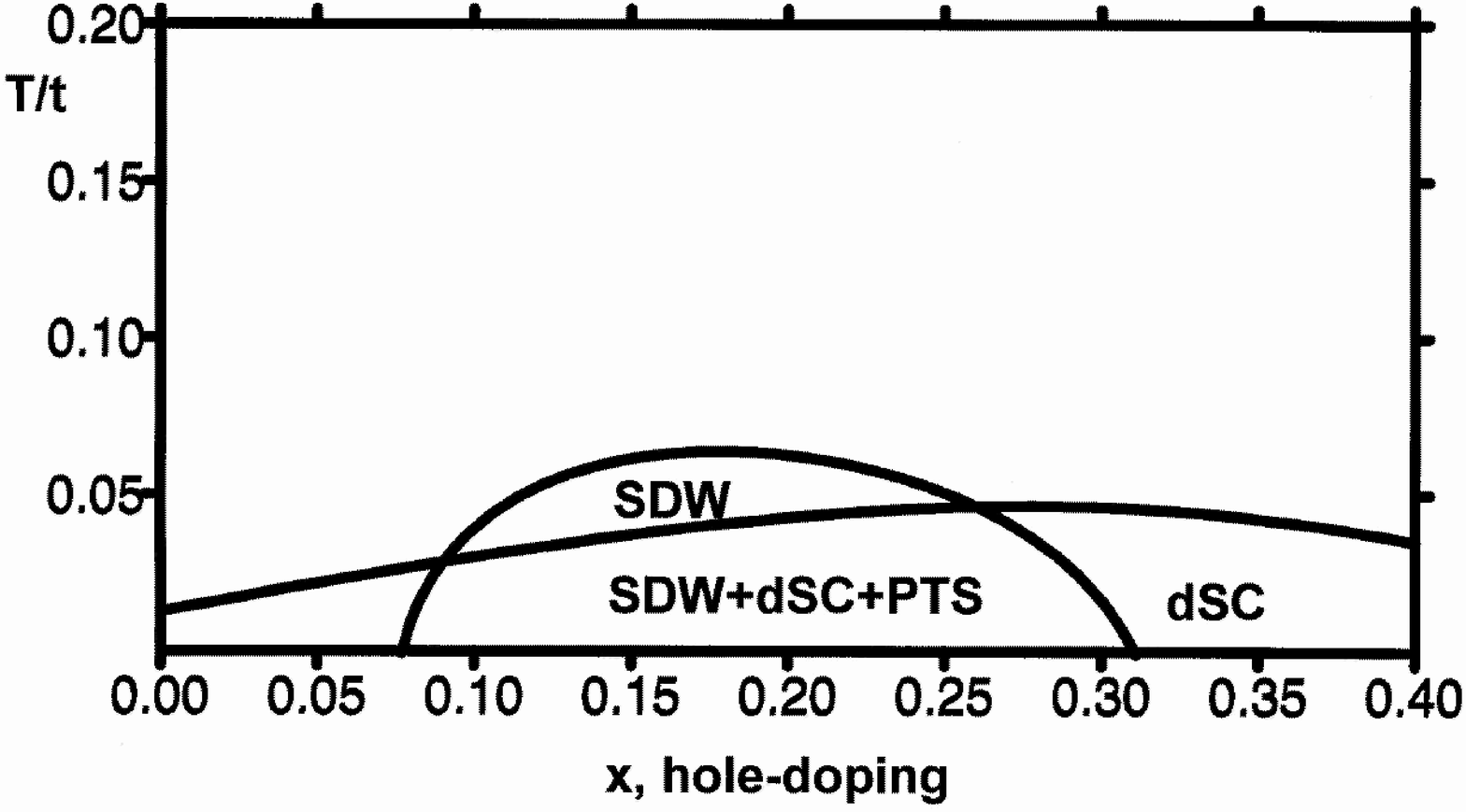}
\caption{Phase diagram for $t/J=1.5$ and $t'/t=-0.3$. T is the temperature. The ordered SDW phase starting from the QCP at $x=x_c$
develops toward the underdoped region. Its size grows with
increasing $J$ and eventually leans out the superconducting phase
as in this figure. SDW and dSC result in a mixed phase where PTS order is also present.}
\label{figure}\end{center}\end{figure}

On the second stage we write a set of coupled equations for different order parameters. We show that the mixed state with both SC and DW order parameters is favorable in the underdoped regime at low temperature (see Fig. \ref{figure}). Moreover the coexistence of DW
and dSC order parameters results in the appearance of a third ordering.
In the case when DW order is of the isotropic SDW type \cite{foot} this third order parameter ($\pi$ triplet superconducting \cite{zhang} (PTS) one ) is
related to the $\pi$ operator introduced in the framework
of the SO(5) theory
 which creates triplet Cooper pairs with total
momentum ${\bf Q}$. 
The resulting phase diagram is shown in Fig. \ref{figure} for the $t-J$ model with realistic for high $T_c$ cuprates parameters.

Notice that the pure DW order develops in the region where the pseudogap behaviour in the normal state is experimentally observed, see e.g. \cite{arpes}. The admixture of SDW order in the dSC state could be responsible for the anomalous spin dynamics observed by neutron scattering inside the superconducting phase. If the dSC order is suppressed, the underdoped regime could be insulating due to the DW order while the overdoped will remain metallic in good agreement with experiment \cite{boebinger} where superconductivity is suppressed by a pulsed magnetic field.

\end{document}